\documentclass{article}

\usepackage[a4paper,margin=1in]{geometry}
\usepackage{graphicx, caption, tabularx, xcolor}
\usepackage{amsmath, amssymb, amsthm, mathtools}
\usepackage[normalem]{ulem}

\usepackage{hyperref}
\usepackage{orcidlink}

\newcommand{\cq}{\color{black}}
\newcommand{\ha}{\color{black}}

\renewcommand{\d}{\mathrm{d}}     

\title{
{Tipping resonance in a chaotically forced ice age model }
}
\author{
Courtney Quinn\footnote{School of Natural Sciences, University of Tasmania, Churchill Avenue, Sandy Bay, 7001, Tasmania, Australia}  $^{\ddagger}$
\orcidlink{0000-0001-5298-5233}, and Hassan Alkhayuon\footnote{School of Mathematical Sciences, University College Cork, Western Road, Cork, T12 XF62, Ireland}  \footnote{Joint lead authors} \, \orcidlink{0000-0001-8117-4907}
}

\date{\today}

\begin{document}
\maketitle



\begin{abstract} 
Many physical systems are forced by external inputs, which can sometimes take the form of chaotic variation. A particular example is found in applications related to weather and climate, where chaotic variation is prevalent across various timescales. If the system in question has multiple attracting solutions for a given range of forcing, rate-induced tipping can be triggered by the chaotic forcing, with the difference in timescales between the forcing and the system acting as a `rate' parameter. In this paper, we explore the interplay between these two timescales in a low-order model of ice age dynamics. The model exhibits bistability between two equilibria in one region of the parameter space and between an equilibrium and a periodic orbit in another region. When chaotic variation of the parameters is allowed within these bistable regions, the solutions of the forced system undergo rate-induced tipping from one attractor to another. Simulations of the forced system show that the timescale of the chaotic forcing induces a resonance-like behaviour, with an optimal timescale at which the likelihood of rate-induced tipping is at its maximum. We combine basin instability theory, finite-time Lyapunov exponents, and linear resonance analysis under periodic forcing to explain this resonance effect.
\end{abstract}

\section{Introduction}
\label{sec:intro}

The concept of tipping has become ubiquitous in the study of natural systems over the past decade,
with the release of the Global Tipping Points Report 2023 \cite{lenton2023global} 
just one indication of the growing interdisciplinary interest in this topic.
From applications viewpoint, ``tipping" is often used to describe when a significant qualitative or quantitative change in a system is observed in time. The definition provided by \cite{lenton2023global} further asserts that the change must be substantial and self-sustaining after passing a threshold, and the shift in the state of the system must remain even if the initial driver is removed (sometimes completely irreversible). 
Earlier definitions, however, were mostly rooted in mathematical theory. 
Examples include loss of stability of an attractor \cite{lenton2008tipping} and threshold behaviour where small changes in initial conditions or parameters cause large and discontinuous changes in long-term outcomes \cite{lamberson2012tipping}. 
In their pioneering paper on tipping in forced nonlinear systems, Ashwin et al. \cite{ashwin2012tipping} differentiate between the mechanisms for tipping from an equilibrium state, based on the properties of the forcing. They coined the terms B-tipping, N-tipping, and R-tipping for bifurcation-, noise-, and rate-induced tipping respectively. {\ha The latter of the three has gotten an increasing amount of attention in recent years
and has been argued as the more significant contributor to environmental shifts in the past century \cite{arnscheidt2022rate, ritchie2023rate, arnscheidt2025rate}.}

For a system with a base state\footnote{By {\em base state}, we mean the desired state of the system prior to any transition taking place.~\cite{wieczorek2023rate}} that is not simply an equilibrium, such as limit cycles, tori, or even chaotic attractors—tipping behaviour can be more complex~\cite{alkhayuon2018rate, keane2020signatures, ashwin2021physical}. For example, nonlinear systems with an oscillatory base state can exhibit Phase-tipping (or P-tipping) when an external input prompts the system to transition to an alternative state, but only from specific phases of the oscillations \cite{alkhayuon2018rate, alkhayuon2021phase, kumar2025pace}. Furthermore, the interplay between the rate of change of external forcing and the phase of the underlying attractor can induce rich dynamic behaviour, such as finite-time unpredictability \cite{chappelle2025rate} or weak tracking \cite{alkhayuon2020weak}.

In the context of forced nonlinear systems, particularly those modelling climate or ecological processes, one can often expect irregularity in external forcing \cite{tyson2016seasonally}. 
The choice in modelling this irregularity varies depending on the dominant signals at play. 
One approach is to use a deterministic chaotic signal in which the characteristics of the chaotic attractor control the nature of the forcing \cite{ashwin2021physical,axelsen2024finite, ashwin2025contrasting, romer2025effect}.
Often, however, the interplay between the timescales of the dynamics and the timescales of the forcing remains generally uncertain. Alkhayuon et al. \cite{alkhayuon2023stochastic} find stochastic resonance associated with 
a climatic switching rate applied to a species population model, namely, there exists an intermediate value of the switching rate
such that the tipping probability from a given attractor reaches a maximum. 
Furthermore, Romer and Ashwin \cite{romer2025effect} examined the behaviour of chaotically forced systems near bifurcation-induced tipping thresholds and identified regions in the parameter space, referred to as the {\em chaotic tipping window}, where tipping can occur only for specific trajectories of chaotic forcing. Outside of this tipping window, the system either tips for all chaotic trajectories or never tips. The number of finite-time trajectories which tip is highly dependent on the forcing timescale and the properties of the chaotic behaviour.

One particular system subject to irregular external forcing on varying timescales is the interaction between global ice cover and atmospheric conditions. The well-known ice age cycles that occurred over the past two million years have been an active area of research for decades.
Frequency of oscillations, asymmetry of the cycles, and transitions between dominant frequencies are among the prominent questions many researchers still attempt to answer \cite{chalk2017causes,verbitsky2018theory,brook2018antarctic,quinn2018mid,herbert2023mid}. 
Given the dominant timescales of interest, it has become a common practice to consider low dimensional models with only a few interacting variables.
This typically includes ice sheet extent and atmospheric CO$_2$, as well as variables representing ocean circulation or other processes on similar timescales. 
One well-known example of this is the collection of Saltzman and Maasch models of the late 80s and early 90s \cite{saltzman1988carbon,maasch1990low,saltzman1990first,saltzman1991first}. These models, while simple in the climate modelling sense, have complex dynamics that are still being explored today (see e.g. \cite{engler2017dynamical,quinn2018mid,ashwin2018chaotic,bosio2023coherence,alexandrov2024role}). 
One open area of research is in the nonautonomous response to forcing of these low-order models. This can take the form of external insolation forcing \cite{maasch1990low,crucifix2012oscillators,quinn2018mid,quinn2019effects,ashwin2018chaotic} or as a variation in the parameters of the model \cite{saltzman1991first}. 
One yet to be considered question is how the timescale of parameter variation affects the solution, and thus the implications of results.

Here we explore this question using
the Saltzman and Maasch 1988 model \cite{saltzman1988carbon}.
This model has previously been shown to operate essentially in a two-dimensional phase space \cite{engler2017dynamical, quinn2019effects}. 
Due to this simplification of dimensionality, the model (or a reduced version) is an ideal candidate for our study. 
To explore the effect of irregular forcing on different timescales, we use a bounded chaotic variation of the parameters but at a different timescale from that of the system. {\cq Importantly, we are interested in an external forcing of parameters which is not affected by the internal dynamics of the system of interest.}
The main aim of this study is to examine the impact of the difference in {\cq forcing} timescale on tipping behaviour between different responses in the model.

The consideration of effects due to the timescale of the chaotic forcing is at its core a ``rate-of-change" problem. 
Thus, we find it convenient to study the tipping behaviour within the framework of R-tipping. One of the important differences between R-tipping and B-tipping is that, with R-tipping, a transition can occur even in parameter regions where the base attractors maintain stability, which can sometimes seem counter-intuitive. To highlight this feature, R-tipping studies tend to restrict analysis to parameter regions where the base attractor of the system is at least locally stable{\cq , i.e. the autonomous attractor does not undergo a bifurcation in that region}. In that context, R-tipping can manifest as a transition between attractors \cite{ritchie2023rate, wieczorek2023rate} or as a large transient excursion from an attractor \cite{o2023rate, vanselow2024rate}.

\begin{figure}
    \centering
    \includegraphics[width=\linewidth]{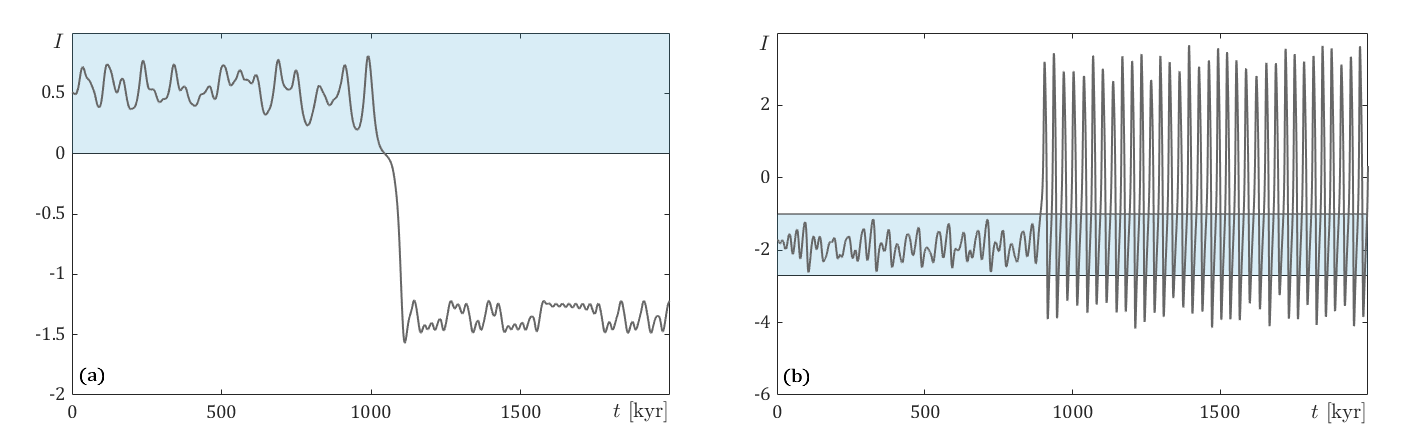}
    \caption{Example trajectories of the transitions of interest: (a) transitioning from a state of anomalously high ice to one of anomalously low ice, and (b) transitioning from a state of anomalously low ice to large amplitude oscillations between low and high ice. The blue shading approximates the 
    {\ha range}
    between the forced response to the base state and an alternative response state.}
    \label{fig:example_trajs}
\end{figure}

In this study, we follow this practice and restrict our analysis to parameter regions where the base state of the system is locally stable for the frozen (unforced) system. Specifically, we are interested in regions of bistability in the frozen system, meaning that, in addition to the stable base state, there exists an alternative state, which can be either a stable equilibrium or a stable periodic orbit for the same parameter values. We choose to consider cases where the base state in the frozen system is an equilibrium. Under chaotic forcing, the transitions of interest will then take one of two forms, as shown in Figure~\ref{fig:example_trajs}. The blue shading represents the {\cq range} for the forced response of the base state. Thus exiting the blue region would represent transitioning to an alternative response. The left plot depicts a transition where both the base state and alternative state are equilibria in the frozen system. In the right plot the alternative state is a periodic orbit in the frozen system. We use $\delta$-close tracking and Finite-time Lyapunov exponents to identify the true boundary between responses in the nonautonomous system. We find that the timescale of the forcing induces a resonance-like effect in the tipping probability for both cases. For the case where the alternative state is a periodic orbit in the frozen system, this resonance behaviour occurs even without basin instability.

The rest of the paper is structured as follows: In Section~\ref{sec:methods}, we present the Saltzman-Maasch nondimensional model~\cite{engler2017dynamical}, its basic bifurcation analysis, our approach to chaotic forcing, and review some tools from basin instability theory~\cite{o2020tipping, wieczorek2023rate}, which we then use to analyse the system. In Section~\ref{sec:region1}, we explore the tipping behaviour from high ice to low ice states, while Section~\ref{sec:region2} examines the tipping behaviour from the low ice state to large amplitude oscillations. Our final discussion and conclusions are presented in Section~\ref{sec:conclusions}.


\section{Methods}
\label{sec:methods}

\subsection{ The two-dimensional Saltzman-Maasch model}
\label{sec:Model}
\begin{figure}
    \centering 
    \includegraphics[width = 1\textwidth]{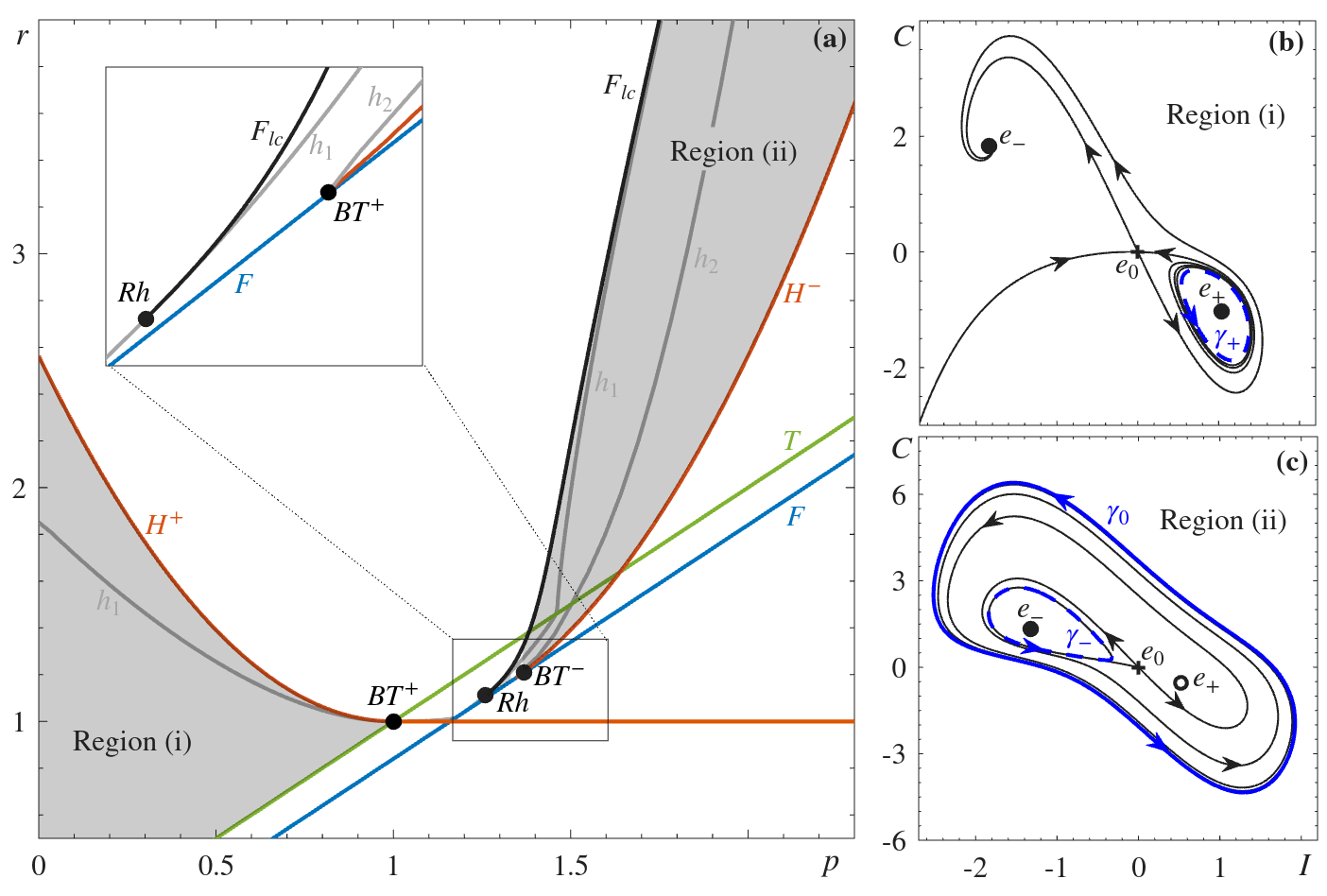}
    \caption{
    Two-parameter bifurcation diagram and phase portraits of System~\eqref{eq:MS_unforced}. (a) The two-parameter $(p,r)$ bifurcation diagram with two shaded regions of bistability. (b) The qualitative phase portrait of the system in region (i), showing two stable equilibria. (c) The qualitative phase portrait of the system in region (ii), showing one stable equilibrium and a large-amplitude stable periodic orbit. 
    The other parameter $s = 0.8$.}
    \label{fig:2par_phase_port}
\end{figure}

{ 
We consider the two dimensional Saltzman-Maasch model for glacial cycles as derived in \cite{engler2017dynamical}: 

\begin{equation}\label{eq:MS_unforced}
    \begin{split}
        \frac{dI}{dt} &=
        -I-C,\\       
        \frac{dC}{dt} &=
        r C + p I + sI^2 - C I^2. 
    \end{split}
\end{equation}
where $I$ represents the nondimensional anomaly in the total global ice mass and $C$ the nondimensional  anomaly in atmospheric CO$_2$. The parameters $r$ and $p$ are nondimensional and
positive 
representing the effective rate of change for $C$ in response to linear changes in $I$ and $C$.
The constant parameter $s$ is an asymmetry parameter which captures the characteristic shape of the ice age cycles; we fix it as $s = 0.8$ as in previous studies \cite{maasch1990low,quinn2018mid}. 



We write
$$
\big(I^0(t), C^0(t)\big) := \phi\big(t,I_0,C_0; p,r\big): \mathbb{R}^5 \to \mathbb{R}^2,
$$
to denote a solution to System~\eqref{eq:MS_unforced} at time $t$, with fixed-in-time parameters $p$ and $r$, and initial conditions $\big(I^0(0), C^0(0)\big) = (I_0,C_0)$.\footnote{The superscript $0$ indicates that we are taking $\epsilon=0$, where $\epsilon t$ is the timescale of variation of the parameters. See Section~\ref{sec:forcing} for more details.}

System \eqref{eq:MS_unforced} has three possible equilibrium solutions: 
\begin{equation}
    \label{MS_equilibria}
    e_+(p,r) = (I_+(p,r), C_+(p,r)), \;\; e_- = (I_-(p,r), C_-(p,r)), \;\; e_0 = (0, 0). 
\end{equation}
where 
$$
I_\pm(p,r) = \frac{-s \pm \sqrt{s^2 +4(r-p)}}{2}, ~\textrm{and}~C_{\pm}(p,r) = -I_{\pm}(p,r). 
$$
We will refer to $e_+$ as the high ice state and $e_-$ the low ice state.  Figure~\ref{fig:2par_phase_port}~(a) shows the stability and bifurcations of the three equilibria in terms of the parameters $r$ and $p$. 
{\ha The bifurcations are organised around two codimension-two bifurcation points. The first point, $BT^-$, is a generic Bogdanov--Takens point \cite{kuznetsov1998elements}: an intersection of a fold curve $F$, a subcritical Hopf curve $H^-$, and a homoclinic bifurcation $h_2$. Note that the fold $F$ and the Hopf curve $H^-$ are bifurcations of the low-ice equilibrium $e_-$.
Furthermore, the system has a large-amplitude stable limit cycle $\gamma_0$, which emerges from a fold of limit cycles $F_{lc}$ together with an unstable limit cycle (not shown). While $\gamma_0$ persists, the unstable limit cycle collides with the saddle equilibrium $e_0$ at the homoclinic bifurcation curve $h_1$ in Region (ii), above the transcritical curve $T$.
Below the transcritical curve, the two equilibria $e_0$ and $e_+$ switch stability and $e_+$ becomes a saddle. Thus, the homoclinic orbit resulting from $h_1$ below $T$ limits to $e_+$ both forward and backward in time.
The fold of limit cycles $F_{lc}$ and $h_1$ intersect tangentially at a codimension-two {\em resonant homoclinic} point $Rh$.

The second organising point, $BT^+$, is a degenerate Bogdanov--Takens point, an intersection of a transcritical bifurcation curve, a subcritical Hopf curve $H^+$, and a homoclinic curve $h_1$. The transcritical curve $T$ and the Hopf curve $H^+$ are bifurcations of the high-ice equilibrium $e_+$.}
For more comprehensive bifurcation analysis of the system we refer to Engler et al. \cite{engler2017dynamical}. 

We are interested in the behaviour of the system when the values of $p$ and $r$ are time-dependent but remain in regions with the same bistability. Of particular interest to us are the shaded regions (i) and (ii) in Figure~\ref{fig:2par_phase_port}~(a):
\begin{itemize}
    \item  
    Region (i) where both the high ice, $e_+$, and the low ice, $e_-$, states are stable. 
    \item 
    Region (ii) where the low ice state $e_-$ is stable, as well as a large amplitude limit cycle $\gamma_0$. 
\end{itemize}
Note that in both regions the local stability of the stable equilibria and limit cycle does not change. In other words, if the parameters $r$ and $p$ change within these two regions there is no possibility of bifurcation-induced transitions from one state to another.
We also point out that in both 
regions the trivial equilibrium $e_0$ is a saddle above the transcritical line. 
{\ha Furthermore, we vary the parameters $r$ and $p$ chaotically on a slower time scale parameterised by $\epsilon > 0$. The details of how we model $r(\epsilon t)$ and $p(\epsilon t)$ are given in Section~\ref{sec:forcing}. This is done throughout this paper apart from Section~\ref{ssec:linear_res}, where we vary the parameters periodically to study the linear resonance effect.
}

}

\subsection{Chaotic external forcing}
\label{sec:forcing}
The motivation for varying $r(\epsilon t)$ and $p(\epsilon t)$ in a chaotic manner relates to the processes they are modelling in the system. The parameter $p(\epsilon t)$ controls the downdraw of atmospheric CO$_2$ into the deep ocean through North Atlantic Deep Water (NADW) formation. On the other hand, the parameter $r(\epsilon t)$ captures the strength of the positive feedback of atmospheric CO$_2$ caused by changes in sea surface temperature, sea ice extent and sea level. The mechanisms driving change in both of these parameters are highly variable and operate on a range of timescales. The formation of NADW is driven by deep ocean circulation and ice sheet dynamics which have dominant variability timescales over the range $10$ yr to $100$ kyr \cite{goosse2010introduction}. A similar large range of timescales from months to millenia exists for sea surface temperature, sea ice extent and sea level changes \cite{taylor1993flickering,king2000middle,rohling2009antarctic,dijkstra2013nonlinear}. Due to the combination of processes captured by these parameters, we choose to model the overall effect by chaotic forcing across the relevant range of timescales. {\cq Chaotic forcing, as opposed to say quasiperiodic forcing, allows for a broad range of frequencies to be captured in a continuum which better describes the uncertainty and nonstationarity in feedback timescales.}

As a paradigm of chaos, we consider the Lorenz 1963 system \cite{lorenz1963deterministic} with time {\ha variable} $\tau=\epsilon t$: 
\begin{equation}\label{eq:Lorenz}
    \begin{split}
        \frac{dx}{d\tau} &=
        \mu(y-x), \\       
        \frac{dy}{d\tau} &=
         x(\rho-z)-y, \\
        \frac{dz}{d\tau} &= xy-\beta z,
    \end{split}
\end{equation}
with the classical chaotic parameters $\mu = 10, \rho = 28$ and $\beta = 8/3$. 
We write 
$$
\big(x(\tau),y(\tau),z(\tau)\big) := \tilde{\phi}(\tau,x_0,y_0,z_0): \mathbb{R}^4 \to \mathbb{R}^3.
$$
to denote a solution to system~\eqref{eq:Lorenz} at time $\tau$, with initial condition $\big(x(0),y(0),z(0)\big) = (x_0,y_0,z_0)$. 


To keep consistency with System \eqref{eq:MS_unforced} time {\ha variable}, we henceforth use $\epsilon t$ in place of $\tau$. Since $x(\tau = \epsilon t)$ is bounded between  $x_\textrm{min}$ and $x_\textrm{max}$, we model the chaotic variation of $p(\epsilon t)$ as

\begin{equation} \label{eq:p_func}
    p(\epsilon t) = (p_2-p_1)\frac{x(\epsilon t) -x_\textrm{min}}{x_\textrm{max} -x_\textrm{min}} + p_1,
\end{equation}
where $p_1$ and $p_2$ are the bounds of variation on $p(\epsilon t)$. As the parameter $r$ is also expected to vary chaotically, we define the bounds of variation $r_1$ and $r_2$. To simplify the tipping analysis, we would like $p(\epsilon t)$ and $r(\epsilon t)$ to vary on a linear parameter path. In other words, we model $r(\epsilon t)$ as a linear function of $p(\epsilon t)$:
\begin{equation} \label{eq:r_func}
r(\epsilon t) = r_1 + \frac{r_2 - r_1}{p_2 - p_1} \big(p(\epsilon t) - p_1\big).
\end{equation}
A parameter path $\mathcal{P}_{ \{p_1,p_2,r_1,r_2\} }$ (or simply $\mathcal{P}$) in the $(p, r)$-plane is then defined as the line segment for $p$ in $[p_1,p_2]$ and $r$ in $[r_1, r_2]$: 
$$ 
\mathcal{P} = 
\left\{ 
(p,r): p \in [p_1, p_2] \, \mathrm{and} \, r = r_1 + \frac{r_2 - r_1}{p_2 - p_1} \big(p - p_1\big)
\right\}.
$$
{\cq In the physical system these parameters might have a nonlinear relationship, however, to connect to the theoretical understanding of time-varying effects we restrict to simple linear paths. There is room for future research to generalise results to a more complex parameter path.}

Note, although the forced system can be augmented to be an autonomous system by combining Eqs (\ref{eq:MS_unforced}), (\ref{eq:Lorenz}), (\ref{eq:p_func}), and (\ref{eq:r_func}), we prefer to look at it as a forced nonautonomous system to make use of the notion of basin instability \cite{o2020tipping} to analyse tipping behaviour.
Finally, we write 
\begin{equation}
\label{eq:forced_sol}
\big(I^\epsilon(t), C^\epsilon(t)\big)
:= \varphi\big(t,I_0,C_0; p(\epsilon t),r(\epsilon t)\big): \mathbb{R}^5 \to \mathbb{R}^2,
\end{equation}
to denote a solution to the full forced system at time $t$, with time-dependent forcing\footnote{The forcing is prescribed by the initial condition of the Lorenz system~\eqref{eq:Lorenz}.} $p(\epsilon t),r(\epsilon t)$ and initial condition $\big(I^\epsilon(0), C^\epsilon(0)\big) = (I_0,C_0)$.

\subsection{Basin instability}
\label{sec:basin_instability}

In this section we revisit the notion of basin instability for equilibria from O'Keeffe and Wieczorek \cite{o2020tipping} and adapt it for system~\eqref{eq:MS_unforced}. 
Basin instability is defined as a property of the autonomous system (i.e. system~\eqref{eq:MS_unforced}), and can serve as a sufficient condition for rate-induced tipping in the nonautonomous system.

We consider system~\eqref{eq:MS_unforced} and define the basin of attraction, $B(e, p,r)$, of a stable equilibrium $e(p,r)$ as the set of all points $(I_0,C_0)$ whose solution converges to $e$:

$$
B(e, p,r) = 
\big\{ 
(I_0, C_0): 
\phi\big(t,I_0,C_0; p,r\big) \to e \, 
{\rm as} 
\, t \to \infty
\big\}.
$$
We also denote the {\em closure} of the basin of attraction by $\overline{B(e; p,r)}$.

Consider a parameter path $\mathcal{P}$ along which the equilibrium $e(p,r)$ is asymptotically stable. We say the equilibrium $e(p,r)$ is {\em basin unstable} on $\mathcal{P}$ if there are two points $(p_1,r_1)$ and $(p_2,r_2)$ in $\mathcal{P}$ such that $e(p_1, r_1)$ is not in the closure of the the basin of attraction of the equilibrium $e(p_2, r_2)$ (see Figure~\ref{fig:BI_MS_model}), i.e.
\begin{equation}\label{eq:BI}
    \begin{split}
        {\rm There \ exist} \, (p_1, r_1) \, &{\rm and} \, (p_2, r_2) \, \in \, \mathcal{P}~{\rm such \ that},\\
        e(p_1, r_1) &\notin \overline{B(e,p_2,r_2)}. 
    \end{split}
\end{equation}

One can show that an equilibrium $e(p,r) = (I_e, C_e)$ that is basin unstable along a parameter path is susceptible to rate-induced tipping.
This implies that there is a parameter forcing $p(\epsilon t),r(\epsilon t)$ that allows the solution of the forced system (\ref{eq:forced_sol}) which starts {\em near} the equilibrium to exit its basin of attraction, and hence tip to an {\em alternative state}. We often refer to the state from which the tipping take place as the {\em base state}, in this case the equilibrium $e$, as opposed to the {\em alternative state} that the system may tip to.

\subsection{Escape, rescue, and tipping events \label{ssec:escape_events}}
In this section we draw a link between the autonomous notions of equilibria, basins of attraction, and basin instability, of the unforced system and the solutions of the forced system.
{\ha
It is important to point out that the solutions of the forced system are not confined to the basins of attraction of the equilibria of the unforced system.
Nevertheless, one can use the unforced system as a quasi-static approximation of the forced system. This means that if the forcing varies extremely slowly, then the behaviour of the solutions of the forced system can be understood using the unforced system.
For faster forcing, on the other hand, the quasi-static approximation fails. 
In this section, we are interested in the instances when this quasi-static approximation fails. Namely, we follow \cite{alkhayuon2023stochastic} and define three events of this quasi-static approximation failure. We call them: {\em escape}, {\em rescue}, and {\em tipping} events. These events will be utilised in Sections~\ref{sec:region1}~and~\ref{sec:region2}.
}
Furthermore, 
we consider two {\ha parameter regions in our discussion}. 
In the first {\ha parameter region}, 
the base state is given by the high ice state $e_+$ and the alternative state is the low ice state $e_-$. 
For this particular case we are interested in parameter paths within region (i) depicted in Figure~\ref{fig:2par_phase_port}(a), where both equilibria $e_-$ and $e_+$ are linearly stable. 

For a {\ha time-dependent parameter point $(p(\epsilon t),r(\epsilon t))$} from region (i),
{\ha we consider the two moving equilibria $e_\pm(\epsilon t)$ that vary with time based on the forcing $(p(\epsilon t),r(\epsilon t))$. Their basins of attraction $B\big(e_-(\epsilon t),p(\epsilon t),r(\epsilon t)\big)$ and $B\big( e_+(\epsilon t),p(\epsilon t),r(\epsilon t)\big)$\footnote{\ha Note that these basins of attraction correspond to the autonomous frozen system and are not nonautonomous dynamical objects.}
are separated by the following borders: 
\begin{itemize}
    \item
    If we choose a point $(p(\epsilon t),r(\epsilon t))$ below the homoclinic curve $h_1$, then the border of the basins of attraction is given by, $W^s(e_0,p(\epsilon t),r(\epsilon t))$, the stable manifold of the saddle $e_0$.
    \item 
    If we choose a point $(p(\epsilon t),r(\epsilon t))$ between $h_1$ and $H_+$, then the border of the basins of attraction is given by, the unstable periodic orbit $\gamma_+(p(\epsilon t),r(\epsilon t))$.
\end{itemize}
} 

We say that a solution of the forced system {\em escapes} the basin of attraction of the equilibrium $e_+(\epsilon t)$ at time $t_{es}$ if the solution lies inside the basin of attraction of $e_+(\epsilon t)$ for a time interval before $t_{es}$ and outside the basin for another time interval after $t_{es}$. 
In other words,
a solution of the forced system (\ref{eq:forced_sol}) escapes the basin of attraction {\ha $B(e_+(\epsilon t),p(\epsilon t),r(\epsilon t))$} at time $t_{es}$ if there exists $t_{es}>0$ and $\delta_1, \delta_2> 0$ such that
{\ha 
\begin{equation*}
    \begin{split}
        \big(I^\epsilon(t), C^\epsilon(t)\big)
        &\in
        B\big(e_+(\epsilon t),p(\epsilon t),r(\epsilon t)\big)\,
        ~\textrm{and}\\
        \big(I^\epsilon(\tilde{t}), C^\epsilon(\tilde{t})\big) &\notin 
        \overline{
        B\big(e_+(\epsilon t),p(\epsilon \tilde{t}),r(\epsilon \tilde{t})\big)
        }
    \end{split}
\end{equation*}
}
for all $t \in (t_{es} - \delta_1, t_{es})$ and $\tilde{t} \in (t_{es}, t_{es} +\delta_2)$.

\begin{figure}
    \centering
    \includegraphics[width=0.7\linewidth]{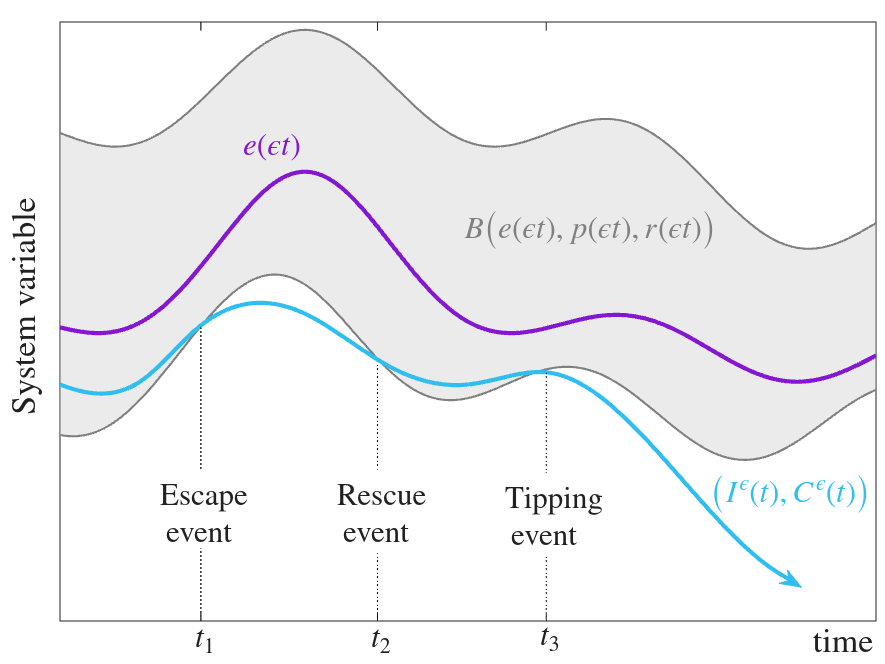}
    \caption{\ha A schematic diagram of escape, rescue, and tipping events. The solution of the forced system (sky blue) escapes the basin of attraction (grey) of the moving equilibrium (dark magenta) of the unforced system at time $t_1$, is rescued at time $t_2$, and escapes again at time $t_3$. No rescue event follows the second escape; thus, the solution tips.}
    \label{fig:esc_res_schem}
\end{figure}

For a solution of the forced system that has already escaped the basin of attraction of $e_+(\epsilon t)$ at time $t_{es}$, we say the solution is {\em rescued} at time $t_{re}$ if the solution lies outside the basin of attraction of $e_+(\epsilon t)$ for the time interval between $t_{es}$ and $t_{re}$, and inside the basin for a time interval after $t_{re}$. 
In other words, 
a solution of the forced system (\ref{eq:forced_sol}) that has already escaped the basin of attraction {\ha $B(e_+(\epsilon t),p(\epsilon t),r(\epsilon t))$} at time $t_{es}$, we say is rescued at time $t_{re}$ if there exists $t_{re}>t_{es}$ and $\delta > 0$ such that:
{\ha 
\begin{equation*}
    \begin{split}
    \big(I^\epsilon(t), C^\epsilon(t)\big)
    &\notin 
    B\big(e_+(\epsilon t),p(\epsilon t),r(\epsilon t)\big)\,
    ~\textrm{and}\\
    \big(I^\epsilon(\tilde{t}), C^\epsilon(\tilde{t})\big) 
    &\in 
    B\big(e_+(\epsilon t),p(\epsilon \tilde{t}),r(\epsilon \tilde{t})\big)
    \end{split}
\end{equation*}
}
for all $ t \in (t_{es}, t_{re})$ and $\tilde{t} \in (t_{re}, t_{re} + \delta)$. Furthermore, we say the solution undergoes 
{\em (rate-induced) tipping} if there is no such $t_{re}$. {\ha We illustrate the escape, rescue and tipping events schematically in Figure~\ref{fig:esc_res_schem}.}

To draw a link between basin instability and escape (tipping) events, we consider a solution of the forced system (\ref{eq:forced_sol}) close to the base state $e_+\big(p(\epsilon t), r(\epsilon t)\big)$ at some initial time $t=t_1$. We suppose the forcing of the system $\big(p(\epsilon t), r(\epsilon t)\big)$ varies on a parameter path $\mathcal{P}$, and the base state $e_+\big(p(\epsilon t), r(\epsilon t)\big)$ is basin unstable on $\mathcal{P}$. 
By the condition for basin instability \eqref{eq:BI} there should be
$(p_1, r_1)$ and $(p_2, r_2)$ in $\mathcal{P}$ such that $e_+(p_1, r_1)$ is not in $\overline{B(e_+,p_2,r_2)}$.
This means that a fast change in the forcing (large $\epsilon$) from $\big( p(\epsilon t_1), r(\epsilon t_1) \big) = (p_1, r_1)$ to $\big( p(\epsilon t_2), r(\epsilon t_2) \big) = (r_2,p_2)$, will lead to an escape event if the solution is close enough to $e_+(p_1, r_1)$ at time $t_1$. 

However, the chaotic variation of the forcing {\ha could imply}
that there exists a time $t_3>t_2$ such that
$$
e_+(p_1, r_1)\in B(e_+,p_3,r_3).
$$
For large enough $\epsilon$ the forcing can change from 
$(p_1, r_1)$ to $(p_2, r_2)$ and then to $(p_3, r_3)$ so quickly that the solution $\big(I^\epsilon(t), C^\epsilon(t) \big)$ escapes and then gets rescued before it diverges away from the time-varying base state. 
Therefore, tipping events are more likely to occur when the forcing changes faster than rate of convergence of the base state, but slower than the rate of divergence from the {\ha basin boundary}. 

In the second {\ha parameter region}, region (ii), 
the base state is given by the low ice state $e_-$ and the alternative state is the large amplitude stable limit cycle $\gamma_0$, see Figure~\ref{fig:2par_phase_port}(c). 
The tipping threshold in this case is the unstable limit cycle $\gamma_-$. 
The discussion of escape, rescue, and tipping events formulated for the previous case can be extended to the second case.

\section{Region (i): 
Tipping from high ice to low ice \label{sec:region1}}

In this section we explore the behaviour of the forced system where variations in the parameters correspond to region (i) in the unforced model (see Figure \ref{fig:2par_phase_port}). In this region both the high ice and low ice states are linearly stable. 
We choose two parameter paths, both with $(p_1,r_1)=(0,1.2)$ but different end points: $(p_2,r_2)=(0.6,1.2)$ only varying in $p$ and ending close to the Hopf bifurcation versus $(p_2,r_2)=(0.5,0.8)$ varying both parameters and ending far from a bifurcation, see Figure~\ref{fig:tipping_e+_MS_model}.\footnote{\ha Note that along both parameter paths, the parameters are varied chaotically as explained in Section~\ref{sec:forcing}.}

\begin{figure}[t]
    \centering \includegraphics[width = 1\textwidth]{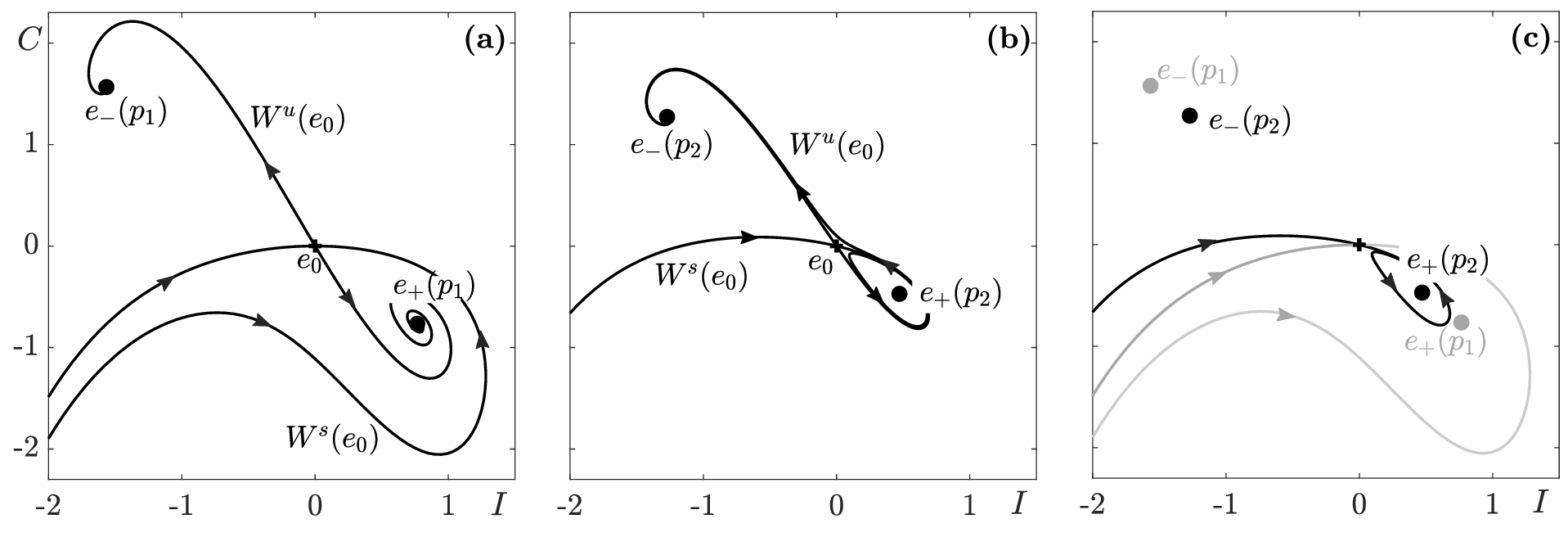}
    \caption{
    {
    Phase portraits of System~\eqref{eq:MS_unforced} at the boundary values of the first parameter path: (a) the parameter values $(p,r) = (p_1,r_1)=(0,1.2)$, (b) the parameter values $(p,r) = (p_2,r_2)=(0.6,1.2)$, and (c) the tipping threshold and equilibria from (b) (in black) with the tipping threshold and equilibria from (a) (in grey), showing that $e_+$ is basin unstable along the path. The other parameter $s=0.8$. 
    }
    }
    \label{fig:BI_MS_model}
\end{figure}

To simplify the following discussion, we first focus on the parameter path which keeps a constant value in $r$. 
The phase portraits corresponding to the boundaries of the $p(\epsilon t)$ variation are shown in Figure \ref{fig:BI_MS_model}. 
We can see from the Figure \ref{fig:BI_MS_model}(c) that there exists basin instability for the high ice state {\ha $e_+(\epsilon t)$} but no basin instability for the low ice state {\ha $e_-(\epsilon t)$}. We will therefore consider instances of escape and rescue events with respect to {\ha $e_+(\epsilon t)$}.

\subsection{Monte Carlo simulations \label{ssec:MC_sim}}

In order to explore the escape and rescue behaviour in this model, we must consider variation in the chaotic forcing. We thus construct a set of Monte Carlo simulations to represent the uncertainty in exact form of the chaotic variation. This is done by varying $x_0$ in the initial conditions to \eqref{eq:Lorenz}. The forcing is thus initialised as follows:
\begin{subequations} \label{eq:L63_IC_x}
    \begin{align}
        x_0 &= x_{\min} + \frac{(x_{\max}-x_{\min})n}{1000}, \quad \textrm{for} \ n = 1,..,1000 , \\
        y_0 &= 0, \\
        z_0 &= 20.
    \end{align}
\end{subequations}
In other words, we take small increments in the initial $x$ value for the Lorenz system as this is the coordinate used in the chaotic parameter variation (\ref{eq:p_func}). The values $x_{\min}$ and $x_{\max}$ are set as a lower bound and an upper bound on the variation observed in the 
$x$-component of the Lorenz attractor; in this case we take $x_{\min}=-19.96$ and $x_{\max}=20$. The initial conditions as defined above result in 1000 simulations. 

As we want to initialise the system
at the high ice state, at the initial forcing 
$p_0$ and $r_0$ calculated from \eqref{eq:p_func} and \eqref{eq:r_func},
the corresponding initial conditions are defined by
\begin{subequations} \label{eq:SM_IC}
    \begin{align}
        I_0 &= -\frac{s}{2} + \frac{1}{2}\sqrt{s^2+4\big(r_0-p_0\big)}, \\
        C_0 &= -I_0.
    \end{align}
\end{subequations}
{\cq Using (\ref{eq:L63_IC_x}) we first solve system (\ref{eq:Lorenz}) for the 1000 simulations. We then input the $x$ trajectories into system (\ref{eq:MS_unforced}) using Eqs. (\ref{eq:p_func}) and (\ref{eq:r_func}) and integrate from the corresponding initial conditions given by (\ref{eq:SM_IC}), taking into account a timescale difference between the two systems.}  Specifically, we take $\epsilon\in[0,1]$ with step size $0.01$. For each value of $\epsilon$ we run {\cq system (\ref{eq:MS_unforced})} for a timespan of $[0,T]$ where $T=250/\epsilon$. This ensures that the trajectories for different values of $\epsilon$ have the same forcing profile. For the singular case when $\epsilon=0$ the systems are uncoupled so we simply let $T=250$. For $\epsilon=1$ this implies we run the model for $T=250$ time units, corresponding to $2500$ kyr, or $2.5$ Myr, which is approximately the length of the Pleistocene epoch.


\subsection{Identification of escape events}

We say that a solution of the forced system undergoes an {\em escape event} if it escapes the basin of attraction of the {\cq moving equilibrium $e_+(\epsilon t)$ in the quasistatic approximation}. 
In this section we use two methods to identify escape events in our Monte Carlo simulations: $\delta$-close tracking \cite{ashwin2017parameter}, and Finite-time Lyapunov exponents (FTLEs) \cite{abarbanel1991}.  

\subsubsection{$\delta$-close tracking} \label{sssec:delta-close}
The notion of $\delta-$close tracking, has been introduced by Ashwin et al.\cite{ashwin2017parameter} in the setting of asymptotically autonomous systems. 
It refers to solutions of the nonautonomous system that stay close to a time varying quasistatic equilibrium of the `frozen' autonomous system. 
Here, we adopt the notion in our chaotically forced system to refer to solutions $\left(I^{\epsilon}(t),C^{\epsilon}(t)\right)$ that evolve within a $\delta$-neighbourhood of the moving base state $e_+ (p(\epsilon t), r(\epsilon t))$.

More precisely, we consider the moving base state $e_+ (p(\epsilon t), r(\epsilon t))$ that is time-dependent. Also, consider a solution of the forced system $\left(I^{\epsilon}(t),C^{\epsilon}(t)\right)$. We say that the solution $\delta$-close tracks the base state $e_+ (p(\epsilon t), r(\epsilon t))$ in the time interval $\mathcal{I}$ if 
$$
\mathrm{dist}\left(
\left(I^{\epsilon}(t),C^{\epsilon}(t)\right),
e_+ (\epsilon t)
\right) < \delta
$$
for all $t \in \mathcal{I}$, where $\mathrm{dist}(X,Y)$ is the Euclidean distance in $\mathbb{R}^2$.
When choosing $\delta$ large enough, but not too large, then one can say that escape events take place when the solution fails to $\delta$-close track the base state for a time interval $\mathcal{J}$. In other words: 

$$
\mathrm{dist}\left(
\left(I^{\epsilon}(t),C^{\epsilon}(t)\right),
e_+(\epsilon t)
\right) > \delta
$$
for some $t \in \mathcal{J}$.\footnote{\ha If a solution of the forced system leaves the $\delta$-neighbourhood of $e_+(\epsilon t)$ and then re-enters it, for a carefully chosen $\delta>0$, then the first exit constitutes an escape event and the re-entry constitutes a rescue event.}

\subsubsection{Finite-time Lyapunov exponents}

We consider the concept of local (in)stability as given by the finite-time Lyapunov exponents of the system (FTLEs). 
In this context the term ``local" refers to nearby points along a solution trajectory. {\cq It is important to note that computation of an asymptotic Lypanunov spectrum assumes the existence of an ergodic measure. For the forcing system (\ref{eq:Lorenz}) it has been shown that a unique invariant measure exists in the chaotic regime used in this study \cite{tucker2002rigorous}. The extended system, in which (\ref{eq:SM_IC}) and (\ref{eq:Lorenz}) are coupled via (\ref{eq:p_func}) and (\ref{eq:r_func}) into a five dimensional autonomous dynamical system, however, has not been shown to have a well-defined attractor with an ergodic measure. Here instead we calculate FTLEs of the nonautonomous solutions $(I^\epsilon(t),C^\epsilon(t))$ to characterise the linear growth and decay over a fixed time window. Where positive FTLEs occur, we expect divergence of nearby trajectories over the chosen window and thus a possibility of escape or tipping events as defined in section \ref{ssec:escape_events}. The use of FTLEs for such an indicator has been demonstrated in both conceptual models (e.g. \cite{quinn2018mid}) and high-dimensional climate models (e.g. \cite{schubert2016dynamical}).} The FTLEs are computed using a QR method as described in Appendix \ref{app:FTLEs}. 


\begin{figure}[h]
    \centering
    \includegraphics[width = 1\textwidth]{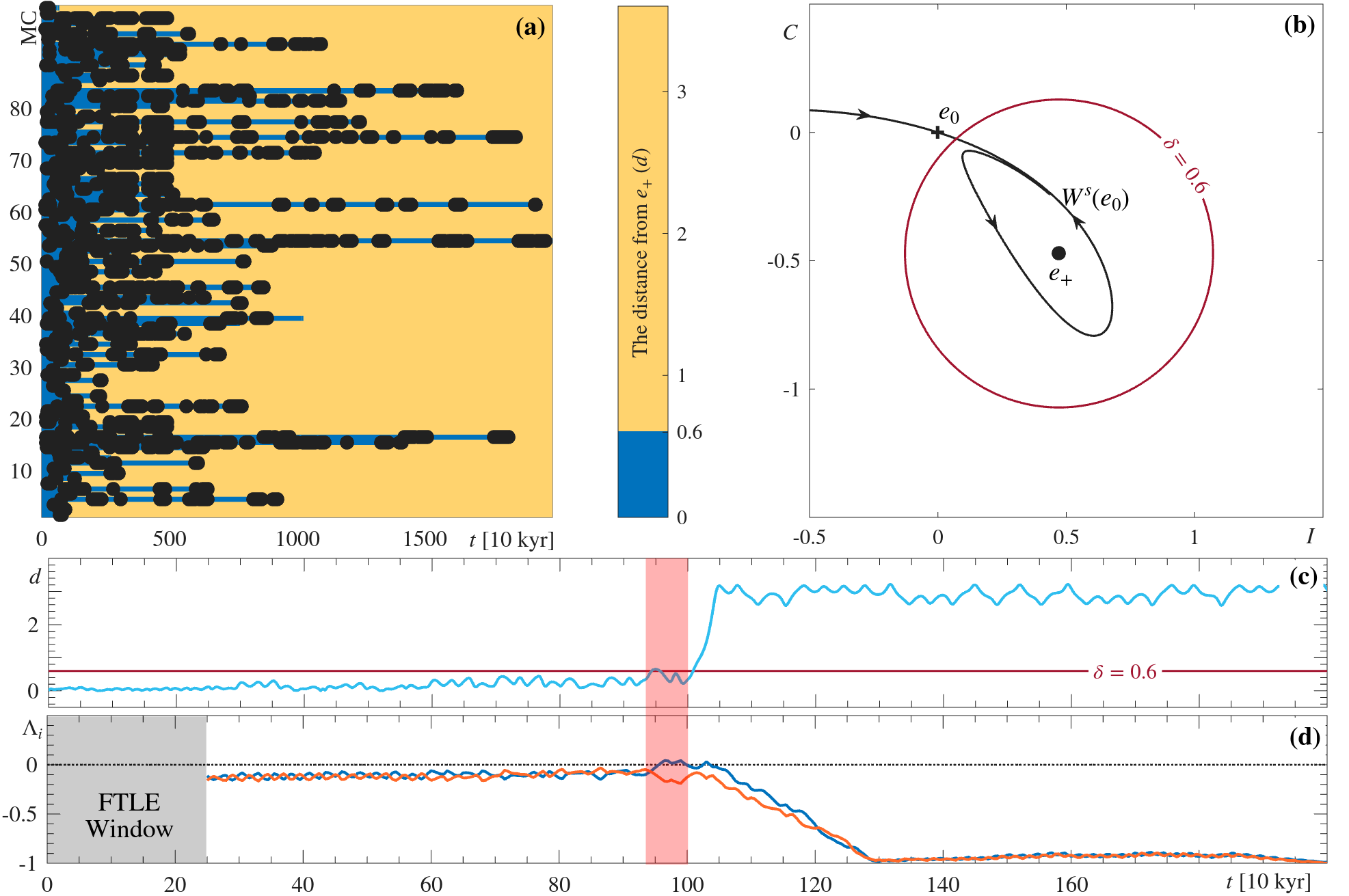}
    \caption{
    { 
    Comparison of distance to the quasistatic equilibrium and Lyapunov exponents for the forced system. (a) Contours of the distance from the quasistatic solution {\cq for a selection of 100 MC simulations}, with black dots denoting where the maximum FTLE is positive. (b) $\delta$-neighbourhood overlaid on the phase plane of the unforced system~\eqref{eq:MS_unforced} with $p=0.6$, $r=1.2$, and $s=0.8$. (c) The distance from the quasistatic equilibrium to one of the solutions that tipped in our Monte Carlo simulation. (d) The FTLEs $\Lambda_1$ (blue) and $\Lambda_2$ (red) along the same solution from (c). The middle {\ha red} vertical shading indicates the time period {\ha from the first crossing to start of deviation from the time-dependent stable manifold of the quasi-static saddle}. All model simulations use $\epsilon=0.2225$.
    }
    }
    \label{fig:dist_Lyaps}
\end{figure}

\subsubsection{Escape events}
\begin{figure}[h]
    \centering
    \includegraphics[width = 0.8\textwidth]{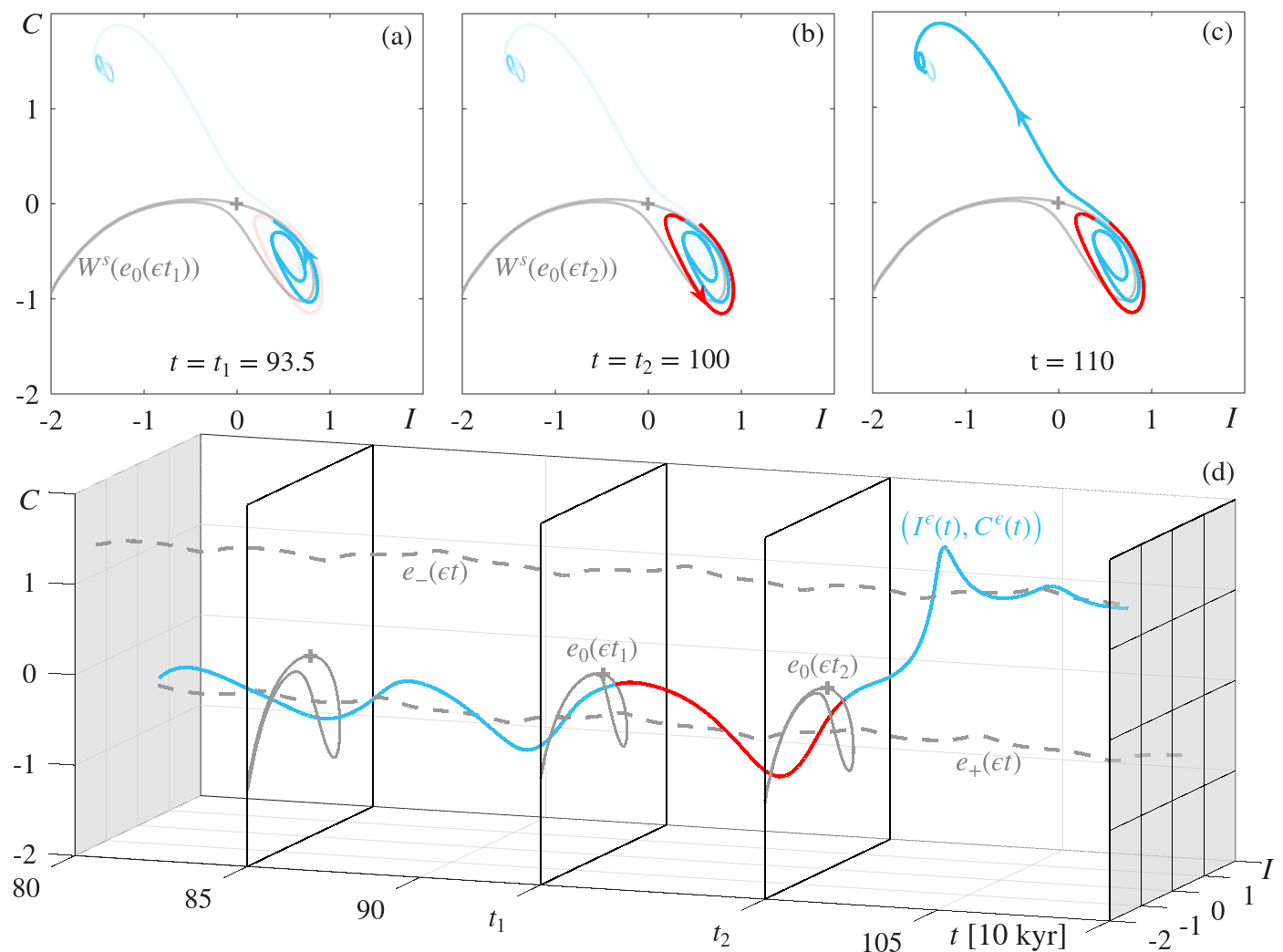}
    \caption{ \ha 
    A solution of the forced system overlaid on the stable manifold (solid grey) of $e_0(\epsilon t)$ at three different snapshots: $t = t_1 = 935$ kyr (a), when the trajectory intersects the stable manifold, $t = t_2 = 1000$ kyr (b), when the trajectory deviates away from the stable manifold, $t = 1100$ kyr (c) after tipping. Subfigure (d) is a 3D visualization of the solution.
    Note that in the interval $t\in[t_1, t_2]$, where the solution is marked in red, the maximum FTLE of the solution attains positive values. This is the same solution used in Figure~\ref{fig:dist_Lyaps} to demonstrate FTLE, and the red interval is the same time interval shaded in red in Figure~\ref{fig:dist_Lyaps}. 
    Simulations are for $\epsilon=0.2225$ and $s=0.8$.
    }
    \label{fig:pp_MC1_pos1}
\end{figure}
We combine both approaches of identifying escape events in order to investigate tipping behaviour.
In the following discussion we fix a forcing {\cq timescale such that there is a dominant period in the chaotic forcing of approximately $28$ kyr; namely} $\epsilon=0.2225$. 

First, from observations of the changing basin of attraction we concluded that $\delta = 0.6$ is a sufficient threshold to capture escape events. 
Figure \ref{fig:dist_Lyaps}(b) shows the phase plane of the unforced system for $p=0.6$ and $r = 1.2$, i.e. the smallest basin of attraction.
Note that, the basin boundary of {\cq $e_+(\epsilon t)$}, i.e. the base state, is given by the stable manifold of the saddle equilibrium {\cq $e_0(\epsilon t)$}. 
The $\delta$-neighbourhood of {\cq $e_+(\epsilon t)$} is overlaid onto the phase plane, showing that for 
some values of $p$, crossing this threshold is indicative of an escape event.

Furthermore, we compute the FTLEs using a sliding window of $T=250$ kyr.\footnote{We chose the same window as in \cite{quinn2018mid}, a sufficiently large window to filter out effects of the average forcing frequency.} 
Figure \ref{fig:dist_Lyaps}(a) shows the occurrence of a positive leading FTLE plotted on top of the distance of each Monte Carlo simulation from its associated quasistatic solution. The color is split such that blue indicates $\delta$-tracking and orange indicates not $\delta$-tracking.
We see that a positive FTLE nearly always precedes a transition, though the time until transition varies.
We plot the distance and FTLEs for an example Monte Carlo simulation in Figures \ref{fig:dist_Lyaps}(c) and \ref{fig:dist_Lyaps}(d), respectively. The first FTLE (blue) indeed increases and becomes positive prior to the trajectory tipping to the alternative quasistatic solution.

To substantiate our conjecture that positive FTLE will 
be observed when the solution of the forced system {\cq experiences an escape or tipping event,} 
we investigate the relationship between a positive FTLE and the crossing of the time-{\cq dependent} stable manifold of the quasi-static saddle. 
{\ha 
In Figure~\ref{fig:pp_MC1_pos1} we show the phase plane behaviour of the same solution of the forced system used to demonstrate FTLE {\cq behaviour} in Figure~\ref{fig:dist_Lyaps}(d). 
The shaded red region in Figure~\ref{fig:dist_Lyaps}(d) corresponds to the red part of the solution in Figure~\ref{fig:pp_MC1_pos1}.
At time $t = t_1 = 935$ kyr the solution of the forced system intersects the time-dependent stable manifold $W^s\big(e_0(\epsilon t)\big)$ of the quasi-static saddle point $e_0(\epsilon t)$, see Figure~\ref{fig:pp_MC1_pos1}(a). 
At time $t = t_2 \approx 1000$ kyr the solution of the forced system starts to deviate from the stable manifold, see Figure~\ref{fig:pp_MC1_pos1}(b). We note that during this time interval $[t_1,\,t_2]$ the leading FTLE attains positive values.
}

\subsection{Resonance in tipping probability}

{The {\cq selection of} 100 Monte Carlo realisations shown in Figure~\ref{fig:dist_Lyaps}(a) suggests that once trajectories escape the $\delta$-neighbourhood, they tip to the alternative low-ice state. 
In all cases visualised in Figure~\ref{fig:dist_Lyaps}(a) the tipping is 
unidirectional, 
by which we mean that the solution can tip from the base state to the alternative state, but once it is sufficiently close to the alternative state, it cannot tip back to the base state.
We visualise the collective tipping behaviour of all 1000 realisations in Figure \ref{fig:eps0.2225_density} through the evolution of the probability density in time. 
Areas with darker shading indicate a high concentration of solutions. We see that some realisations tip quite early and around half of realisations tip by $t=325$. Nearly all realisations eventually tip under this value of $\epsilon$.

We are thus interested in tipping probability when varying the timescale of the forcing. 
{ We consider two different parameter paths as depicted in Figure~\ref{fig:tipping_e+_MS_model}(a). Note that both enter the region of basin instability, one approaching the Hopf bifurcation curve (blue) and the other sufficiently far from a bifurcation curve (red).}
We conduct Monte Carlo simulations for a range of $\epsilon\in[0,1]$ as described in Section \ref{ssec:MC_sim}. For each $\epsilon$ we record the number of solutions which at the final time  of our simulation ($t=250/\epsilon$) are not $\delta$-close to the base state.
We express this number as a probability for each value of $\epsilon$ and show the resulting  curves in Figures~\ref{fig:tipping_e+_MS_model}(b-c).
We notice that in both cases there is a minimum value for $\epsilon$ below which no tipping occurs. This is analogous to the critical rate \cite{ritchie2023rate} in the case of rate-induced tipping. 
In Figure~\ref{fig:tipping_e+_MS_model}(b), this value is around $\epsilon=0.09$, while in Figure~\ref{fig:tipping_e+_MS_model}(c) it is $\epsilon=0.12$.
Additionally, for $\epsilon>0.6$ the 
tipping probability becomes very minimal in both cases. 
This is expected in that for a fast forcing timescale the system response is dominated by the forcing rather than internal dynamics. A more peculiar feature of the tipping probability is the existence of multiple peaks with variable widths. This is more pronounced in the parameter path which approaches the Hopf bifurcation, noting the multiple peaks in Figure~\ref{fig:tipping_e+_MS_model}(b).

We hypothesise that these peaks result from a resonance effect occurring at specific timescales. 
One can examine this effect through the lens of {\em coherence resonance} \cite{palenzuela2001coherence,pisarchik2023coherence}.  
Coherence resonance is a phenomenon in excitable systems with noise, whereby, if one defines a coherence measure for the oscillatory response of the system, this measure attains its maximum at an optimal level of noise. Furthermore, in our system, the chaotic forcing plays a role similar to {\ha the}
noise. 
In our case, we are interested in forcing timescales that causes the solution to tip; hence, we consider tipping probability as our ``coherence measure".  
In the following section, we analyse these resonances in terms of the forcing frequency and the frequency of solutions spiralling towards the base state $e_+$.  

\begin{figure}
    \centering 
    \includegraphics[width = \textwidth]{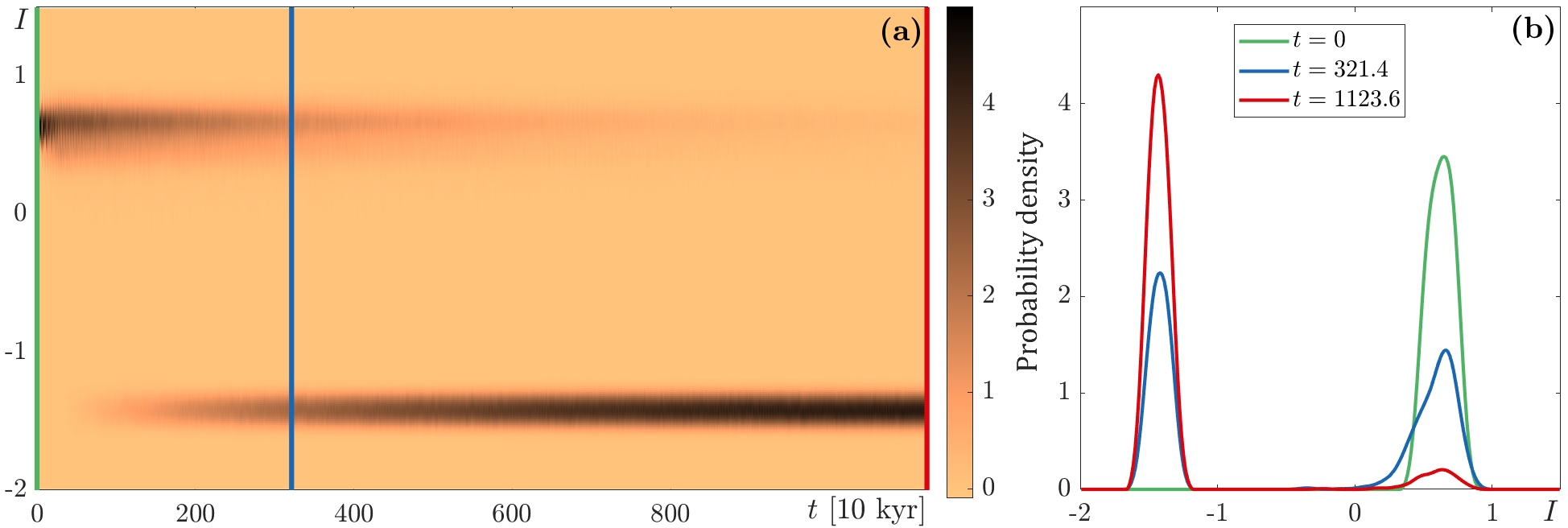}
    \caption{ 
    Probability density estimate of 1000 Monte Carlo simulations of the forced system along the (blue in Figure~\ref{fig:tipping_e+_MS_model}) parameter path with $p_1=0$, $p_2=0.6$, $r{1,2}=1.2$, $s=0.8$, and $\epsilon=0.2225$: (a) evaluated at every time step, and (b) snapshots of the probability density estimate at $t=0$, $t=321.4$, and $t=1123.6$. See the animated version (\texttt{densityMCs\_movie.mp4}) in the electronic supplementary material.
    }
    \label{fig:eps0.2225_density}
\end{figure}

\begin{figure}
    \centering \includegraphics[width = 1\textwidth]{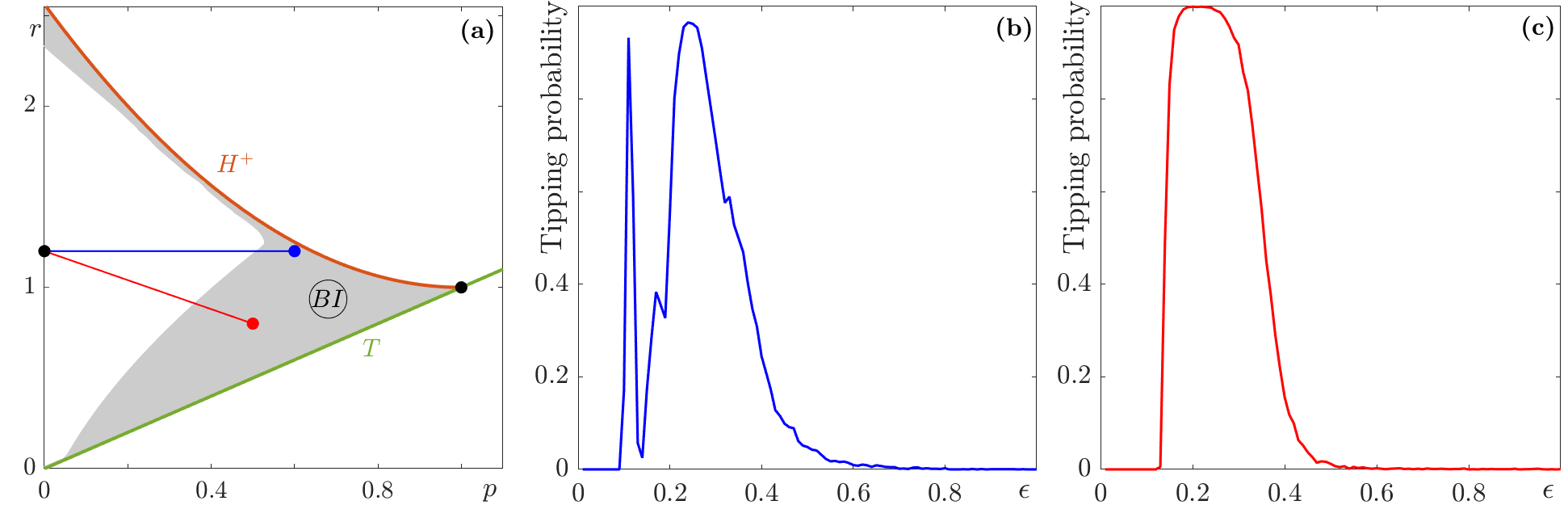}
    \caption{
    Tipping probability from $e_+$ to $e_-$ along two parameter paths. (a) Two-parameter bifurcation diagram with basin instability region of $e_+$ from the black dot $(p_1,r_1) = (0,1.2)$. Both parameter paths, (blue) $(p_2,r_2) = (0.6,1.2)$ and (red) $(p_2,r_2) = (0.5,0.8)$ cross the basin instability regions. (b-c) Tipping probability along the two paths 
    as described in Section \ref{ssec:MC_sim}, the colours corresponds to parameter path shown in (a). The other parameter $s=0.8$.}
    \label{fig:tipping_e+_MS_model}
\end{figure}

\subsubsection{Optimal timescale for resonance} \label{sec:optimal_timescale}

If we think of resonance in its most basic sense, then for a given $\epsilon$ there exists a rational relationship between the internal frequency of our system and the {\cq dominant} forcing frequency, i.e.
\begin{equation}
    \frac{\omega_{SM}}{\epsilon\omega_L}=\frac{m}{n}, \quad m,n\in\mathbb{N}.
\end{equation}
Here, $\omega_{SM}$ is the internal {\cq angular} frequency and $\omega_L$ is the forcing {\cq angular} frequency {\cq in their respective phase spaces}. This can also be rearranged for $\epsilon$ to find timescales associated with particular resonances,
\begin{equation}\label{eq:resonance_cond}
    \epsilon = \frac{\omega_{SM}}{\omega_L} \frac{n}{m}, \quad m,n\in\mathbb{N}.
\end{equation}

We estimate
the dominant forcing frequency 
by the complex eigenvalues of the nontrivial equilibria of the Lorenz system, $\hat{e}_\pm$. The nontrivial equilibria $\hat{e}_\pm$ are given as
\begin{equation}
    (x_{\hat{e}_\pm},y_{\hat{e}_\pm},z_{\hat{e}_\pm}) = (\pm\sqrt{\beta(\rho-1)},\pm\sqrt{\beta(\rho-1)},\rho-1).
\end{equation}

For the canonical values of the Lorenz model used in this study, the equilibrium has eigenvalues $\hat{\lambda}_\pm \approx 0.094 \pm 10.1945i$ and $\hat{\lambda}_s \approx -13.8546$. Note these eigenvalues are the same for $\hat{e}_-$ due to the symmetry of the system. Thus we are interested in resonances involving what we will denote the Lorenz frequency, $\omega_L = 10.1945$.
In Table \ref{tab:eigs_eplus} we list the eigenvalues of {\cq $e_+(\epsilon t)$} under varying $p$ in line with the chaotic variation associated with Figure \ref{fig:tipping_e+_MS_model}. We also include the eigenvalues of the {\cq autonomous} equilibrium at its nearby Hopf bifurcation ($p_{H^+}=0.642229$).

 

In considering resonance with the various {\cq autonomous} internal frequencies, we found that the local peaks seen in Figure \ref{fig:tipping_e+_MS_model} (including the dominant one) correspond to resonances with the frequency of the Hopf bifurcation. This kind of resonance has been seen in systems that exhibit a delayed Hopf bifurcation and are subject to periodic forcing \cite{baer1989slow,su1997effects}. Here we would like to emphasise that our parameter path is bounded away from the Hopf point $p_{H^+}$, i.e. the system never crosses the bifurcation, even though the effect of resonance is clear.  
%
%
If we consider $\omega_{H^+}=0.8705$ as given in Table \ref{tab:eigs_eplus} and the internal frequency of the Lorenz forcing $\omega_L$, then we find the local peaks in tipping probability occur near the values of $\epsilon$ that correspond to $2:1$, $3:1$, and $4:3$ resonances. 

\begin{table}
    \centering
    \begin{tabular}{|c||c|c|c|c|}
        \hline
        $p$ & 0 & 0.3 & 0.6 & $p_{H^+}$ \\[5pt]
        $\omega_{SM}$ & 1.3227 & 1.1343 & 0.9069 & 0.8705 \\[5pt]
        $\mu_{SM}$ &  -0.1935 & -0.0982 & -0.0113 & 0
        \\ \hline 
    \end{tabular}
    \caption{Eigenvalues of $e_+$ under varying $p$ ($r=1.2$, $s=0.8$). Here $\omega_{SM}$=Im($\lambda_\pm$), $\mu_{SM}$=Re($\lambda_\pm$) and $p_{H^+}=0.642229$.}
    \label{tab:eigs_eplus}
\end{table}

\subsection{Resonance in periodic forcing} \label{ssec:linear_res}

To further explore the relationship between the dominant frequency of the chaotic forcing and the frequency of the subcritical Hopf bifurcation, we consider a simple periodic variation of $p$. This will be defined by 
\begin{equation} \label{eq:p_func_sin}
    p(\epsilon t) = (p_2-p_1)\frac{\sin(\epsilon\omega_Lt+\phi)+1}{2} + p_1, \quad \phi\in[-\pi,\pi].
\end{equation}
Note that the parameter $\phi$ allows us to more directly connect our results to the concept of \textit{phase tipping}, namely the tipping of certain phases of a periodic attractor under a parameter shift with critical rate \cite{alkhayuon2018rate}. We show this behaviour for an individual choice of $\epsilon$ in Appendix \ref{app:PhaseTipping}.

In the following subsections we first consider the linear response to a small amplitude periodic variation around a value of $p$, namely $p_0$. We then compare the linear response amplitude across different values of $p_0$ and $\epsilon$ to the tipping probability of the full nonlinear system forced by (\ref{eq:p_func_sin}).

\subsubsection{Linear response to periodic forcing}
Here we consider the response of the system under a small-amplitude periodic variation in $p$. Namely, we take
\begin{equation}
    p(\epsilon t) = p_0+{\cq\eta}\sin(\epsilon \omega_{L}t).
\end{equation}
Here ${\cq\eta} \ll 1$ and $\omega_L$ is the dominant frequency of the Lorenz attractor as introduced in Section \ref{sec:optimal_timescale}. We expand around $e_+$ in orders of ${\cq\eta}$:
\begin{subequations}
\begin{align}
    I(t) &= I_++{\cq\eta} I_1(t)+O({\cq\eta}^2), \\
    C(t) &= C_++{\cq\eta} C_1(t)+O({\cq\eta}^2), 
\end{align}
\end{subequations}
Substituting this expansion into system \ref{eq:MS_unforced} results in
\begin{subequations}
\begin{align}
    {\cq\eta} \frac{\d I_1}{\d t} =& -I_+-{\cq\eta} I_1-C_+-{\cq\eta} C_1+O({\cq\eta}^2), \\
    {\cq\eta} \frac{\d C_1}{\d t} =& \;rC_++{\cq\eta} rC_1+p_0I_++{\cq\eta} p_0I_1+{\cq\eta} I_+\sin(\epsilon\omega_Lt)+ \nonumber \\ 
    & sI_+^2+{\cq 2{\cq\eta} sI_1I_+-C_+I_+^2-2{\cq\eta} C_+I_+I_1-{\cq\eta} I_+^2C_1}+O({\cq\eta}^2). 
\end{align}
\end{subequations}
We see that at $O(1)$ $I_+$ and $C_+$ satisfy the conditions for an equilibrium of system (\ref{eq:MS_unforced}) as expected. At $O({\cq\eta})$ we obtain the linearized system of equations
\begin{subequations}
    \begin{align}
        \frac{\d I_1}{\d t} =& -I_1-C_1, \label{eq:lin_I}\\
        \frac{\d C_1}{\d t} =& \;rC_1+ p_0I_1+ I_+\sin(\epsilon\omega_Lt)+2sI_+I_1- 2C_+I_+I_1-I_+^2C_1. \label{eq:lin_C} 
    \end{align}
\end{subequations}
Rearranging (\ref{eq:lin_I}) we have $C_1=-I_1-\dot{I}_1$. Substituting this expression for $C_1$ in (\ref{eq:lin_C}) gives an inhomogeneous second-order linear ODE for $I_1$:
\begin{equation}
    \frac{\d^2I_1}{\d t^2}+(1-r+I_+^2)\frac{\d I_1}{\d t}+(p_0-r+2sI_++3I_+^2)I_1=-I_+\sin(\epsilon\omega_L t).
\end{equation}
Solving in the typical manner gives linear response
\begin{equation}
    I_1(t) = \frac{I_+(\epsilon\omega_L(1-r+I_+^2)\cos(\epsilon\omega_Lt)+(p_0-r+2sI_++3I_+^2)\sin(\epsilon\omega_Lt))}{(\epsilon^2\omega_L^2-p_0+r-2sI_+-3I_+^2)^2+\epsilon^2\omega_L^2(1-r+I_+^2)^2},
\end{equation}
or simply
\begin{equation}
    I_1(t)=k_c(\epsilon)\cos(\epsilon\omega_Lt)+k_s(\epsilon)\sin(\epsilon\omega_Lt).
\end{equation}
The amplitude of the oscillation can then be calculated as a function of $\epsilon$ through
\begin{equation} \label{eq:linear_response}
    A(\epsilon)=\sqrt{k_c(\epsilon)^2+k_s(\epsilon)^2}.
\end{equation}
}

\subsubsection{Comparison of linear and nonlinear behaviour}
\begin{figure}
    \centering
    \includegraphics[width=0.6\linewidth]{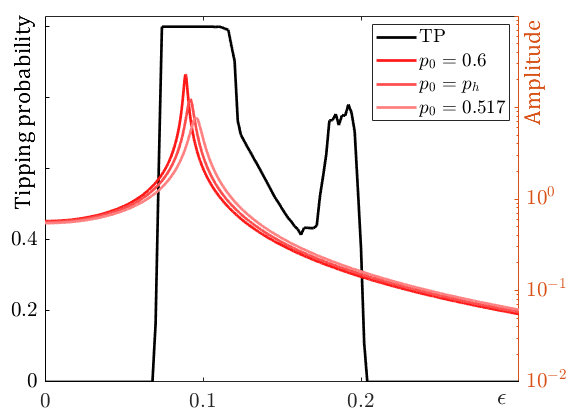}
    \caption{Black line shows tipping probability to $e_-$ for the forced system when starting from $e_+$ under sinousoidal forcing with varying phase (1000 realisations) as defined by Equation (\ref{eq:p_func_sin}). Red lines indicate the amplitude of the linear response as derived in Section \ref{ssec:linear_res} for different values of $p_0$.}
    \label{fig:ptip_eqforc_sin_wres}
\end{figure}

We take 1000 evenly distributed phases around the unit circle and perform Monte Carlo simulations of system (\ref{eq:MS_unforced}) under parameter forcing defined by (\ref{eq:p_func_sin}) for a range of $\epsilon$ values with $p_1$ and $p_2$ as defined previously. An example of the behaviour for $\epsilon=0.16$ is shown in Appendix \ref{app:PhaseTipping}. We again compute the tipping probability for $\epsilon\in[0,1]$ in a similar manner to Figure \ref{fig:tipping_e+_MS_model}. The results are plotted in Figure \ref{fig:ptip_eqforc_sin_wres} with the linear amplitude response for various $p_0$ given by Eq.~(\ref{eq:linear_response}) overlaid. We see that when the response amplitude surpasses $\mathcal{O}(1)$, the likelihood of tipping rapidly increases. The second relative maximum observed in the tipping probability for $\epsilon\in(0.177,0.194)$ appears to be the range for superharmonic resonance of the linear peaks shown.

\section{Region (ii):  
Tipping from low ice to large amplitude oscillations \label{sec:region2}}

We now turn our attention to the transition between a low ice state, $e_-$, and large amplitude oscillations, $\gamma_0$. When modulated with the chaotic variability, this manifests as smaller scale glacial-interglacial cycles on the timescale of the forcing to irregular occurrences of slow glaciation with much faster deglaciation (see Fig.~\ref{fig:example_trajs}(b)). Such a shift between these two responses has been suggested as a mechanism for the Mid-Pleistocene Transition \cite{maasch1990low,quinn2018mid}.

\subsection{Initialising tipping probability calculation \label{ssec:MC_sim2}}

Again we consider a set of Monte Carlo simulations to represent the uncertainty in exact form of the chaotic variation. We vary $x_0$ in the initial conditions to \eqref{eq:Lorenz} and initialise simulations as described in Section \ref{ssec:MC_sim}. The main difference is in the initial conditions of (\ref{eq:MS_unforced}), now taken at the low ice state and thus defined as
\begin{subequations} \label{eq:SM_IC2}
    \begin{align}
        I_0 &= -\frac{s}{2} - \frac{1}{2}\sqrt{s^2+4\big(r_0-p_0\big)}, \\
        C_0 &= -I_0.
    \end{align}
\end{subequations}

We also must set a different criterion for determining whether a simulation has tipped. As displayed in Figure \ref{fig:2par_phase_port}c, the distance from the equilibrium value varies widely along the periodic orbit and large excursions from the base state may occur without a transition being observed. In order to capture true transitions, we use the $\delta$-tracking method described in Section \ref{sssec:delta-close} with a very large $\delta=5$. As long as the system has transitioned with enough time to complete one cycle around the periodic orbit, the tipping will be identified.

\subsection{Tipping resonance on three parameter paths}

\begin{figure}
    \centering
    \includegraphics[width=0.8\linewidth]{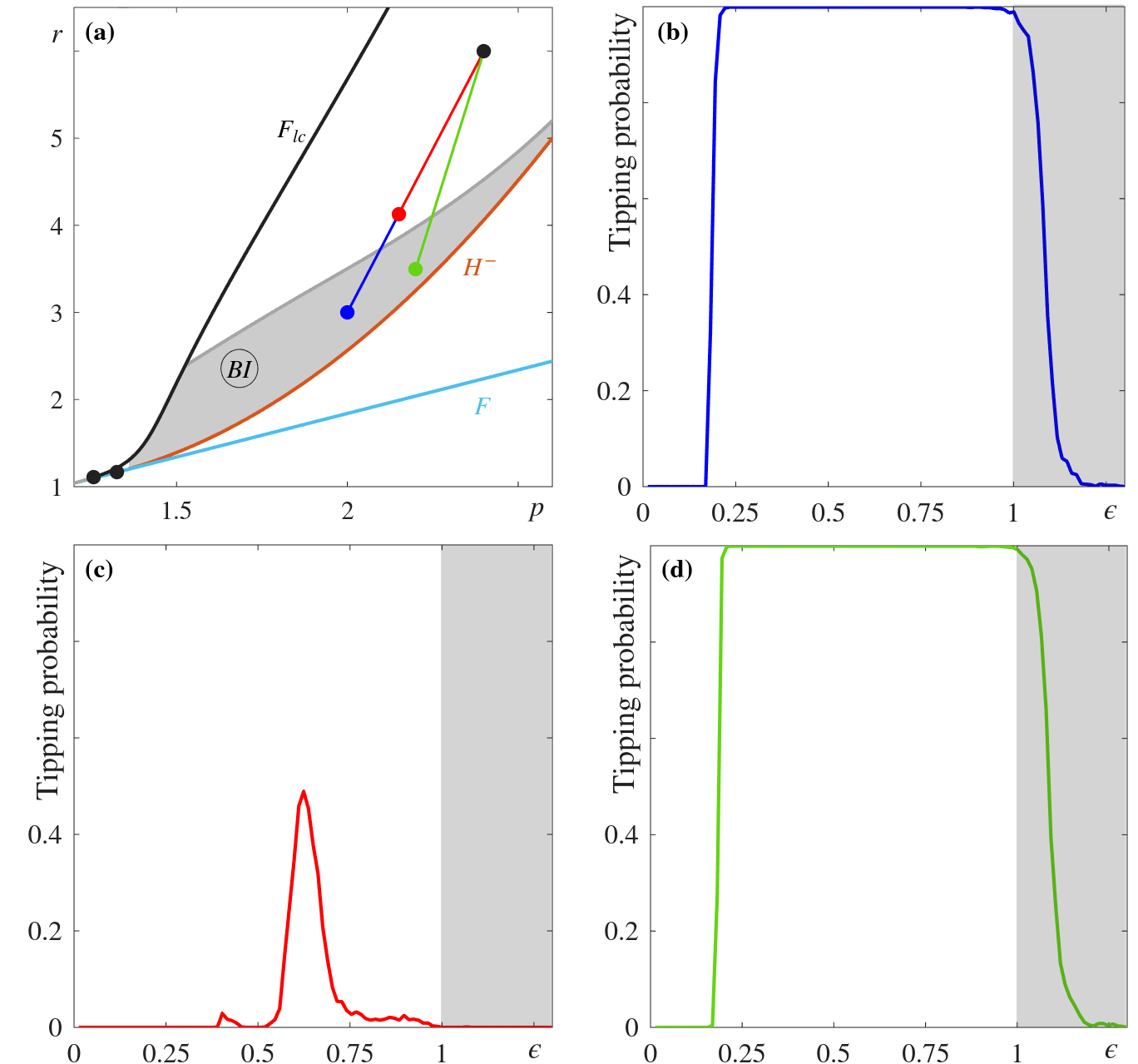}[h]
    \caption{
    Tipping probability from $e_-$ to the large amplitude periodic orbit $\gamma_0$ along three parameter paths. (a) Two-parameter bifurcation diagram with basin instability region for $e_-$ from the black dot $(p_1,r_1) = (2.4,6)$. Two of the parameter paths (blue) $(p_2,r_2) = (2,3)$ and (green) $(p_2,r_2) = (2.2,3.5)$ cross the basin instability region; the (red) $(p_2,r_2) = (2.15,4.125)$ path does not cross the basin instability region. (b-d) Tipping probability along the three paths 
    as described in Section \ref{ssec:MC_sim}, the colours corresponds to parameter path shown in (a). The other parameter $s=0.8$.
    } 
    \label{fig:BI_TP_emtoPO_3paths}
\end{figure}

Here we consider three parameter paths as shown in Figure~\ref{fig:BI_TP_emtoPO_3paths}. All paths have base state $e_-$ defined for $(p_1,r_1)=(2.4,6)$ but each has a different end point for the parameter path: $(p_2,r_2)=(2,3)$ shown in blue, $(p_2,r_2)=(2.15,4.125)$ shown in red, and $(p_2,r_2)=(2.2,3.5)$ shown in green. 
The base state $e_-$ is basin unstable on the blue and the green paths, but it is basin stable on the red path. 
In other words, the base state at any given point on the path is contained in its basin of attraction at all other points on the path.
This is demonstrated in the movie \texttt{BI\_movie.mp4} provided in the electronic supplement.  
Nevertheless, that does not prevent R-tipping from taking place.
We would like to point out that R-tipping from a base state that is basin stable on a parameter path has been previously observed in a canonical example by Xie \cite[Section~4.4]{xie2020rate}.

We observe very similar behaviour for the two basin unstable paths. Namely, there is a minimum epsilon value such that tipping can occur ($\epsilon \approx 0.196$). 
After surpassing this timescale, the system rapidly approaches full tipping (\emph{i.e.} all 1000 realisations of the chaotic forcing experience tipping to the periodic orbit). If only considering behaviour up to $\epsilon=1$, or the chaotic forcing on the same timescale as the system, it would appear that tipping always occurs for $\epsilon \gtrsim 0.2$. However, for illustrative purposes we extend the transition probability curve to $\epsilon>1$ and we observe a decay of tipping for larger $\epsilon$ values. This is in line with the results of \cite{romer2025effect} which found that as the timescale of chaotic forcing gets very large, the behaviour limits to stochastic forcing. This means the behaviour of all realisations can be approximated by the average, and given much of the parameter path remains far from a bifurcation we do not expect tipping. In other words, we indeed observe a resonance in the tipping probability as a function of $\epsilon$, albeit for a very large range of timescales. Note that the rate of decay in the tipping probability beyond $\epsilon=1$ will scale with the time step of the numerical integration method.

The case of the stable parameter path, i.e. the red path, 
shows quite different behaviour. First, the forcing timescale at which tipping is first observed is higher, ($\epsilon \approx 0.39$). There appears to be much more obvious resonance behaviour in the tipping probability, with a clear maximum around $\epsilon \approx 0.624$. The tipping probability decays much before $\epsilon=1$ and we observe no tipping for $\epsilon>1$. Another notable feature is that the tipping probability never reaches one. Due to the finite-time nature of our simulations, this manifests similar to partial-tipping or phase tipping from non-equilibrium attractors. In the asymptotic limit all realisations will eventually tip. However, from an applications standpoint it is highly relevant to know the finite-time probability of tipping. A discussion of phase tipping in the context of periodic forcing is presented in Appendix \ref{app:PhaseTipping}. Here we are suggesting that the finite-time realisation of the forcing induces a similar partial-tipping behaviour dependent on the segment of the attractor visited by the forcing profile. See connections to work \cite{romer2025effect}.

\section{Conclusions}
\label{sec:conclusions}
{\ha 
In this paper, we examined rate-induced tipping in the two-dimensional Saltzman-Maasch model for glacial cycles \cite{engler2017dynamical} under chaotic forcing. 
To model the chaotic forcing, we used the Lorenz 1963 system \cite{lorenz1963deterministic} and assumed that it evolved on a different timescale $\epsilon t$, whereas the Saltzman--Maasch model evolved on $t$. 
We imposed parametric forcing along parameter paths bounded away from any dangerous bifurcations \cite{thompson1994safe} and along which the Saltzman-Maasch model remains bistable at each point. 
Guided by basin instability analysis \cite{wieczorek2023rate}, we identified paths that induce tipping from the high- to low-ice state and from the low-ice state to large-amplitude oscillations purely due to timescale differences.
By sampling different initial conditions of the Lorenz system, we found that the tipping probability is maximised at intermediate values of the timescale parameter $\epsilon$, resembling coherence resonance \cite{palenzuela2001coherence}. 
We then explained this resonance behaviour by applying concepts from basin instability theory and linear resonance analysis under periodic forcing.
}

Our results have important implications for both the mathematical community and those interested in modelling systems subject to irregular external forcing.
Motivated by applications in Earth systems and climate science, researchers are now taking more interest in
chaotically forced bistable systems \cite{ashwin2021physical, mehling2024limits, romer2025effect, axelsen2024finite}, aiming to understand the intricate mechanisms underlying critical transitions in these systems. We hope that our study contributes to this effort by highlighting the crucial interplay between the timescale of the forcing and the timescale of the underlying system. 
This interplay may allow a bistable system to undergo a transition even without passing through a classical dangerous bifurcation.
Examples of some critical Earth system tipping elements with bistability include the Atlantic Meridional Overturning Circulation \cite{alkhayuon2018rate}, the Greenland Ice Sheet \cite{robinson2012multistability}, and the Amazon rainforest \cite{hirota2011global}. All of these elements are subject to chaotic external forcing on a large range of timescales.
By demonstrating that such a transition that is purely induced by the timescale of the forcing 
can occur in the well-studied two-dimensional Saltzman-Maasch model for glacial cycles, we aim to draw attention to this phenomenon and encourage others to explore models that may be subject to irregular forcing across a varying range of timescales.


\section*{Code accessibility}
{
All codes used in this work are available in the GitHub repository:\\ \texttt{github.com/hassanalkhayuon/chaotic\_bistability}
}

\section*{Funding and acknowledgements}
C.Q. is supported by the Australian Research Council Discovery Early Career Researcher Award project number DE250101025. We thank Larry Forbes for some insightful discussions which aided our analytical results. We would also like to thank Ma\l gorzata O'Reilly for her advice regarding the calculation of time-dependent probability density functions. We are grateful as well to Peter Ashwin, Raphael Romer, Theodore Vo, Bernd Krauskopf, Hinke Osinga, Paul Ritchie, {\ha Peter Ditlevsen, and an anonymous referee for insightful and helpful comments at various stages of this work.}

\appendix
\begin{section}{Calculating finite time Lyapunov exponents}\label{app:FTLEs}
    \paragraph{Tangent space mapping.}

Suppose that we have a given solution to System \eqref{eq:MS_unforced}, $\phi(t,I_0,C_0;p,r)$. For brevity we set $\phi(t)\equiv \phi(t,I_0,C_0;p,r)$. A small perturbation in phase space $\delta\phi(t)\in\mathbb{R}^2$ evolves in the tangent space as
\begin{equation}
    \delta\phi(t+\tau)=\mathrm{exp}\big(J(\phi(t+\tau))\tau\big)\delta\phi(t),
\end{equation}

where $J(\cdot)$ is the Jacobian of the right-hand side of \eqref{eq:MS_unforced}. Note that as the system is nonlinear this will have dependence on the solution trajectory itself.

We consider the mapping of the tangent space flow,
\begin{equation}
    \delta\phi(t)_{n+1}=M_n\delta\phi(t)_n,
\end{equation}
and define $M_n=\mathrm{e}^{J(\phi(t)_n)h}$ where $\phi(t)_n=\phi(t+nh)$, $\delta\phi(t)_n=\delta\phi(t+nh)$ and $h$ is a small step size. 

\paragraph{QR method.}

We use a continuous QR algorithm for computing the Lyapunov exponents (see \cite{dieci1997}). We initialise an arbitrary orthogonal $Q_0=\mathbf{I}_2$, noting that $\mathbf{I}_{2}$ is the $2\times 2$ identity matrix.  We define interatively $Q_i$ and $R_i$ by the QR decomposition of $M_iQ_{i-1}$ (MATLAB command qr($M_iQ_{i-1},0$)),
\begin{equation}
Q_iR_i = M_iQ_{i-1},
\end{equation}
which produces a square $2\times 2$ upper triangular matrix $R_i$ with eigenvalues $R_{i,jj}>0$ ($j = 1,2$).  We store $R_i$ for each timestep.  After $N$ timesteps we have the relation
\begin{equation}
Q_NR_N...R_1 = M_N...M_1Q_0.
\end{equation}
The infinite Lyapunov exponents can then be approximated by
\begin{equation}
\lambda_j = \frac{1}{N}\sum_{i=0}^{N}\ln R_{i,jj}.
\end{equation} 
For the FTLEs we truncate the above sum after $W$ timesteps, and view how this truncation changes in time, i.e.
\begin{equation}
\lambda_{j,n} = \frac{1}{hW}\sum_{i=n-W}^{n}\ln R_{i,jj}.
\end{equation} 
Here $W = w/h$, where $w$ is the desired time window length.  Note that $n$ initialises at time $W$.

\end{section}

\begin{section}{Phase tipping under periodic forcing}\label{app:PhaseTipping}

Here we show the results for the Monte Carlo simulations under periodic forcing as described in Section~\ref{ssec:linear_res} when $\epsilon=0.16$. The probability density in time is visualised in Figure \ref{fig:density_sin_eps16}. We see that some of the realisations transition while others maintain tracking the base state. In order to understand this 
through the lens of 
phase tipping, Figure \ref{fig:phasetip_sin_eps16} displays the initial condition of each realisation in three different ways: the initial phase along the unit circle, the value of $(I_0,C_0)$ in the phase plane, and the value $p(0)$ for each phase $\phi$. The initial condition is then coloured as to whether the trajectory tips or tracks the base state. We see there is a nontrivial relationship between the initial phase and tipping behaviour.

\begin{figure}
    \centering
    \includegraphics[width=\linewidth]{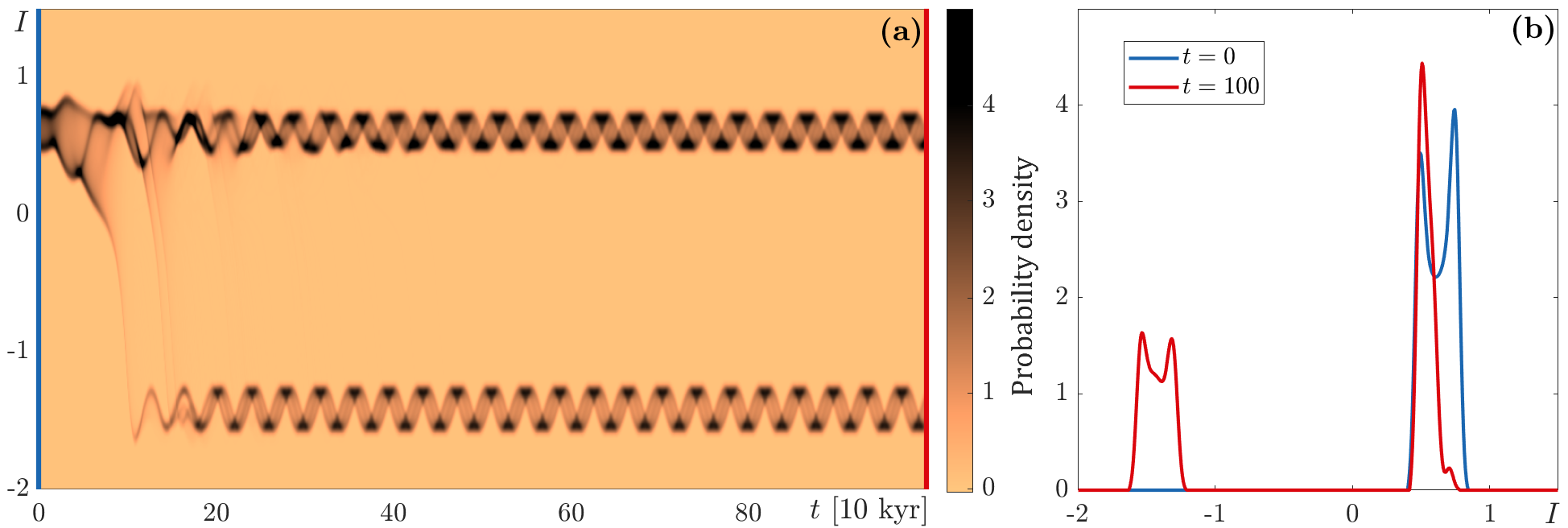}
    \caption{
    Probability density estimate of 1000 Monte Carlo simulations of the forced system with periodic forcing~\eqref{eq:p_func_sin} with $\epsilon=0.16$: (a) evaluated at every time step, and (b) snapshots of the probability density estimate at $t=0$ and $t=100$.
    }
    \label{fig:density_sin_eps16}
\end{figure}

\begin{figure}
    \centering
    \includegraphics[width=0.8\linewidth]{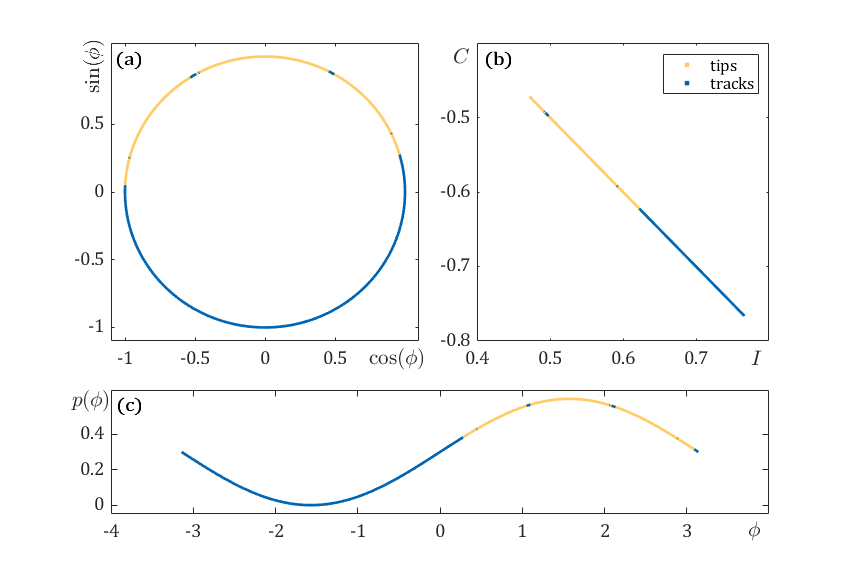}
    \caption{
    Phase tipping behaviour of 1000 Monte Carlo simulations of the forced system with periodic forcing~\eqref{eq:p_func_sin}, where $\epsilon=0.16$. (a) Location of phase along the unit circle, (b) base state in the phase plane, and (c) the value of the parameter $p$ as a function of the phase of the forcing. In all plots, gold signifies tipping while blue signifies tracking. The other parameter values are $r=1.2$ and $s=0.8$.
    }
    \label{fig:phasetip_sin_eps16}
\end{figure}
\end{section}


\bibliographystyle{unsrt}
\bibliography{biblio.bib}

\end{document}